\documentclass[12pt]{iopart}

\usepackage{graphicx}
\usepackage[labelformat=simple]{subcaption}
\usepackage{hyperref}
\usepackage{booktabs} 
\usepackage{algorithm}
\usepackage{algpseudocode}
\usepackage{academicons}

\usepackage{svg}
\newcommand{\orcid}[1]{\href{https://orcid.org/#1}{\includegraphics[width=8pt]{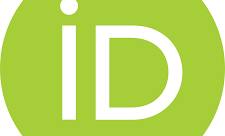}}}

\usepackage{graphicx}
\usepackage{dcolumn}
\usepackage{bm}
\usepackage[square,sort,comma,numbers]{natbib}
\usepackage{multirow}
\usepackage{xcolor}
\usepackage{amssymb} 
\expandafter\let\csname equation*\endcsname\relax
\expandafter\let\csname endequation*\endcsname\relax
\usepackage{amsmath} 
\usepackage{xspace}
\usepackage[english]{babel}
\usepackage{hyperref}
\usepackage{float}
\usepackage{upgreek}
\usepackage{siunitx}

\newcommand{\ptmiss}{\ensuremath{p_\mathrm{T}^\text{miss}}\xspace}

\newcommand{\hlsfml}{\texttt{hls4ml}\xspace}
\def\AA{\textrm{A\kern -1.3ex\raisebox{0.6ex}{$^\circ$}}\kern 0.4ex}
\newcommand{\DN}{\textsc{DistillNet}}

\begin{document}

\title{\boldmath Distilling particle knowledge for fast reconstruction at high-energy physics experiments}





\author{A. Bal$^1$, T. Brandes$^1$, F. Iemmi$^2$, M. Klute$^1$, B. Maier$^{1,3}$, V. Mikuni$^4$, T. K. \AA rrestad$^5$}

\address{$^1$ Karlsruhe Institute of Technology (KIT), Karlsruhe, Germany}
\address{$^2$ Institute of High Energy Physics (IHEP), Beijing, China}
\address{$^3$ I-X, Imperial College, London, United Kingdom}
\address{$^4$ National Energy Research Scientific Computing Center (NERSC), Berkeley Lab, Berkeley, USA}
\address{$^5$ Eidgen\"ossische Technische Hochschule Z\"urich (ETH), Zurich, Switzerland}

\ead{benedikt.maier@cern.ch}


\begin{abstract}
Knowledge distillation is a form of model compression that allows artificial neural networks of different sizes to learn from one another. Its main application is the compactification of large deep neural networks to free up computational resources, in particular on edge devices. In this article, we consider proton-proton collisions at the High-Luminosity LHC (HL-LHC) and demonstrate a successful knowledge transfer from an event-level graph neural network (GNN) to a particle-level small deep neural network (DNN). Our algorithm, \DN, is a DNN that is trained to learn about the provenance of particles, as provided by the soft labels that are the GNN outputs, to predict whether or not a particle originates from the primary interaction vertex. The results indicate that for this problem, which is one of the main challenges at the HL-LHC, there is minimal loss during the transfer of knowledge to the small student network, while improving significantly the computational resource needs compared to the teacher. This is demonstrated for the distilled student network on a CPU, as well as for a quantized and pruned student network deployed on a field-programmable gate array. Our study proves that knowledge transfer between networks of different complexity can be used for fast artificial intelligence (AI) in high-energy physics that improves the expressiveness of observables over non-AI-based reconstruction algorithms. Such an approach can become essential at the HL-LHC experiments, e.g., to comply with the resource budget of their trigger stages. 
\end{abstract}

\vspace{2pc}





\maketitle
\flushbottom

\section{Introduction}
\label{sec:intro}

The Large Hadron Collider (LHC) produces complex data that requires high-energy physics (HEP) experiments like ATLAS~\cite{ATLAS:2008xda} or CMS~\cite{CMS:2008xjf} to employ sophisticated models and simulation tools in order to understand the fundamental constituents of the universe and their interactions. Over the past years, state-of-the-art machine learning solutions have found their way into the data analysis workflows because of their ability to map the data into a set of latent features, a so-called representation, with favorable properties for tasks such as reconstruction, classification, or regression. Best representation learning capabilities in HEP data have been demonstrated by large-scale models based on graph neural networks and transformers, for instance, for the canonical task of jet characterization~\cite{Mikuni:2020wpr,Mikuni:2021pou,Qu:2022mxj,Gong:2022lye,He:2023cfc} and for the reconstruction of the entire collision event~\cite{ExaTrkX:2020nyf,Pata:2021oez,Maier:2021ymx,Li:2022omf}. These models, however, often have millions of parameters, and a forward pass can produce latencies of several seconds per event. This makes them a realistic option only in the offline processing of  large recorded and simulated datasets, where massive parallelization is possible on distributed computing resources.

Especially for the online selection steps of the analysis chain, but in extreme cases also for the offline processing of data, such models can quickly become impractical or even impossible to deploy. In an effort to make data processing more efficient, researchers in other fields have turned to knowledge distillation (KD), first introduced in~\cite{hinton2015distilling} as a form of model compression. It allows the extraction of valuable insights from complex models and represents them in a more compact form. A typical application of KD is the transfer of knowledge from a large, complicated model, referred to as \emph{teacher model}, to a simpler neural network, called \emph{student model}, which fits the resource budget of the specific computing environment consisting of processors and/or co-processors.

A generic loss function used for the KD process can be formulated as 
\begin{align}
\mathcal{L} = \alpha\,{S}(\hat{y},y)+\beta\,{S}(\hat{y},y')+\gamma\,{T}(y,y'),
\label{eq:kd}
\end{align}
with $\hat{y}$ being the ground truth, $y$ the output of the teacher (e.g., class logits or regressed quantities), and $y'$ the student output. Accordingly, ${S}(\hat{y},y)$ is a supervised criterion for aligning the teacher output with the ground truth, similarly for the student output through ${S}(\hat{y},y')$. The core knowledge transfer term ${T}(y,y')$ makes the student mimic the behavior of the teacher. The latter term can, for instance, be a mean squared error between the student and teacher outputs, or the Kullback-Leibler divergence~\cite{kullback1951information} between their distributions. The parameters $\alpha$, $\beta$, and $\gamma$ are relative, tunable weights.

A first successful demonstration of KD in high-energy physics for streamlining data analysis and reducing computational requirements has been performed in~\cite{compression}, where a large convolutional neural network (CNN) was compressed into a smaller CNN able to identify muons by energy deposits in resistive plate chambers. A second application improves the jet tagging capability of lightweight networks~\cite{Liu:2023dio}.

In our specific case,  we consider the problem of pileup mitigation, which is one of the main challenges of future data taking periods at the LHC. It is the identification of particles erroneously attributed to the collision of interest when colliding bunches containing $10^{11}$ of protons. Being one of the key challenges at the Large Hadron Collider (LHC), earlier studies have demonstrated that GNNs and attention mechanisms can quite accurately deduce the provenance of the reconstructed particle~\cite{Maier:2021ymx,Li:2022omf,PhysRevD.108.096003}. They achieve their performance by considering the neighborhood of a particle and learning from shower shapes a \emph{soft label} that indicates if the particle comes from a pileup collision (soft label $\sim$0) or the primary collision (soft label $\sim$1). These labels can then be used to construct precise high-level observables describing the primary collision. However, this comes at a significant computational cost due to inference times of O(s), rendering the application of such algorithms and their practicality quite limited. 

In this article, we study if the soft labels provided by a large, pre-trained graph neural network (GNN) as featured in~\cite{PhysRevD.108.096003} can be used as regression targets for a simple deep neural network, \DN, to achieve good physics performance at lower computational cost. The fact the teacher is pre-trained implies that $\alpha$ of ~\autoref{eq:kd} is set to 0. Likewise, the parameter $\beta$ is 0, because hard labels $\hat{y}$ do not exist, as discussed in the following. The term $T$ denoting the transfer of knowledge between the algorithms is given by the mean absolute error loss,
\begin{align}\label{eq:MAE_loss}
T(y,y')\equiv\mathcal{L}_{\mathrm{MAE}}=\frac{1}{N}\sum_{i=1}^{N} |y_i-y'_i|\;,
\end{align}
where $N$ denotes the batch size given by the number of particles, $y_i$ is the soft truth label from the GNN for particle $i$, and $y'_i$ is the prediction by \DN~for the same particle.

\begin{figure*}[t]
    \centering
    \includegraphics[trim=0cm 0.0cm 0cm 0cm, clip, width=0.825\textwidth]{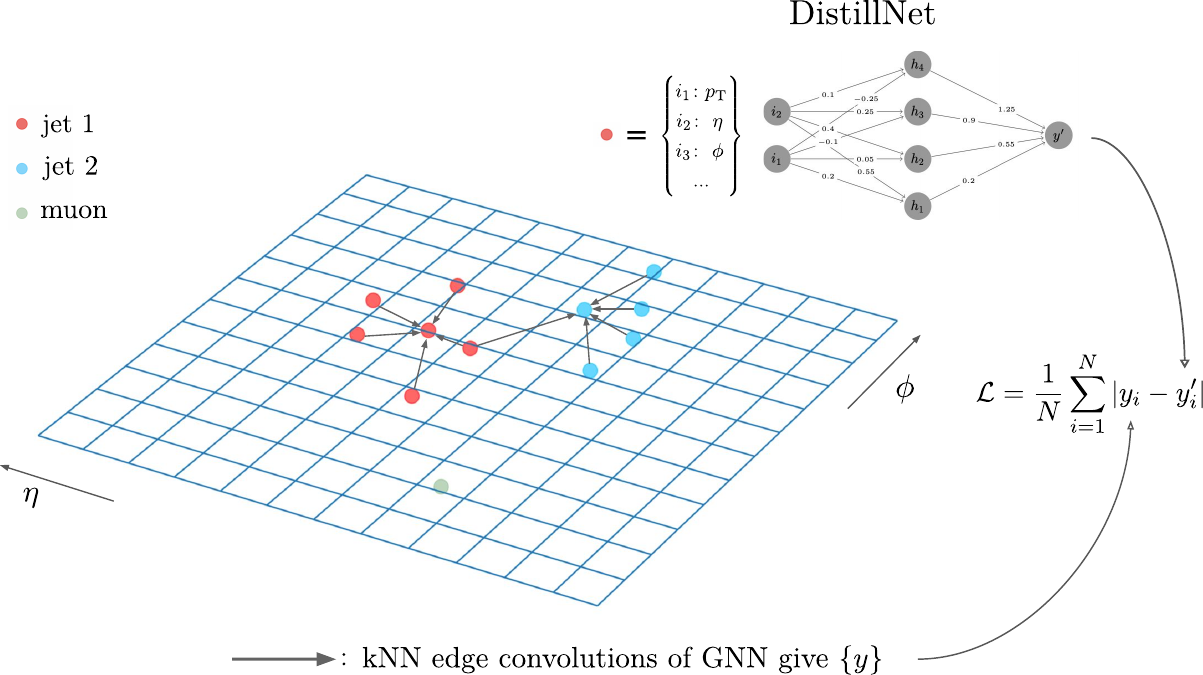}
    \caption{The process of obtaining soft labels $y$ with a GNN making use of kNN edge convolutions (in this illustration $k=5$, in the actual GNN $k=20$). A label $y_i$ reflects the GNN knowledge about the provenance of particle $i$. \DN regresses this label with a simple DNN architecture such that the mean absolute error between $\{y\}$ and $\{y'\}$ is minimized.}
    \label{fig:sketch}
\end{figure*}

What makes the problem of pileup mitigation an ideal use case for studying knowledge transfer from a GNN to a DNN is the intractability of \emph{hard labels} (called $\hat{y}$ in ~\autoref{eq:kd}) for the particles that a DNN could be trained on directly. This is due to the complexities of the simulation, the detector response, and the reconstruction, all of which break the unique correspondence between reconstructed particle candidates and incident truth particles, rendering any plausible ground truth that could be obtained through a human annotation process insufficiently sharp (see~\cite{PhysRevD.108.096003} for a more detailed discussion). Moreover, no truth labels are available in real data. The GNN in~\cite{PhysRevD.108.096003} sidesteps this problem with a self-supervised training procedure, comparing samples with different pileup levels to deduce soft labels for the reconstructed particle candidates. Through a series of nearest-neighbor graph convolutions, the GNN output is a set of soft labels $\{y\}$ enriched with message-passing information, representing the full knowledge of the GNN about the origin of the particle candidates. Being the only available labels, our aim is to use a per-particle DNN to regress these targets, using the information about particle neighborhood  implicitly through the targets that are the GNN output $\{y\}$. That is, we distill the GNN knowledge about the particles into a small DNN. 
For illustration purposes, ~\autoref{fig:sketch} depicts the process of obtaining the GNN and DNN outputs. 

The following sections contain a description of the GNN and, respectively, DNN, as well as a comparison of their performance in terms of physics and resource requirements. We present two different variants of the network, one optimized for performance that can be deployed on a CPU, and one tailored to fit on field programmable gate arrays (FPGA) to operate at real-time in the first online selection stage (``Level-1 trigger'') of the HL-LHC experiments. 

\section{Algorithm}
In the spirit of a simple student network, \DN~is a feed-forward regression neural network tasked to regress a per-particle pileup weight based on the input particle features. It is composed of an input layer with 16 nodes, two hidden layers with 128 and 64 neurons, respectively, and an output layer with a single neuron. The connections between the input layer and the first and second hidden layers use the Rectified Linear Unit (ReLU) activation function. The output neuron is finally transformed with a sigmoid activation function. In addition, a batch normalization layer is incorporated between the second hidden layer and the output node. After conducting initial tests employing 128 nodes for both hidden layers, a subsequent hyperparameter scan revealed that reducing the size of the second hidden layer results in more consistent outcomes, and that increasing the layer size beyond 128 nodes does not yield a significant improvement. The algorithm description of the model is described in Algorithm~\autoref{alg:training}.

The training data are composed of individual particles sampled from W$(\ell\upnu)+$jets events. The events have been generated with the \textsc{Pythia} v8.244 event generator~\cite{pythia8,Corke_2011} and reconstructed with the \textsc{Delphes} v3.4.3pre01 simulation~\cite{delphes} of the Phase-II CMS detector that includes a highly granular forward calorimeter and extended tracker coverage compared to the CMS Run-2 detector. This detector simulation gives us so called E-Flow particle candidates that are the result of combining and linking the information from the various subdetectors. The center-of-mass energy of the collisions is 14\,TeV. On average, 140 pileup collisions have been added to the primary collision, with tails reaching 200 simultaneous collisions. 
\DN~is trained on 15 million particles for 40 epochs. Each particle is characterized by 16 features, a description of which can be found in~\autoref{tab:Distilinput}. These features are identical to the ones used in the training of the teacher GNN.
\begin{algorithm}
    \caption{\textsc{DistillNet} Training}\label{alg:training}
    \begin{algorithmic}
    \Require Data set $\mathcal{D}$, teacher model GNN$_\theta$, distillation model DNN$_\phi$
    \While{not converged}
        \State $\textbf{x}\sim\mathcal{D}$ \Comment{Sample event-level inputs from data set}
        \State $\mathbf{y} = \mathrm{GNN}_\theta(\mathbf{x})$ \Comment{Evaluate teacher GNN to get soft labels}
        \State $\mathbf{x}\rightarrow x$, $\mathbf{y}\rightarrow y$ \Comment{Flatten particle inputs and soft label outputs}
        \State $y' = \mathrm{DNN}_\phi(x)$ \Comment{Evaluate the distillation model on single particles}
        \State $L_\phi = \mathcal{L}_{\mathrm{MAE}}(y,y')$ \Comment{Calculate the distillation loss}
        \State $\phi \leftarrow \phi - \gamma\nabla_\phi L_\phi$ \Comment{Update distillation model parameters}
    \EndWhile
    \end{algorithmic}
\end{algorithm}
After training to convergence, the physics performance of \DN~is evaluated on an independent set of particles from W$(\ell\upnu)+$jets events and on particles from another physics process, the production of a top quark pair with dileptonic decays of the top quarks. We find consistent behavior for the different processes and therefore only present results for top quark pair production.

\renewcommand{\arraystretch}{1.2}
\begin{table}[h!]

    \centering
    \caption{Particle features used as input for \DN.}
    
    \begin{tabular}{ll}
    \toprule
        Variable & Description\\
\hline

        $\eta$              & Particle pseudorapidity                                          \\

        $\phi$              & Particle azimuthal angle                                         \\

        $\log{p_\text{T}}$  & Log of the particle transverse momentum                          \\

        $\log{E}$           & Log of the particle energy                                       \\

        $Q$                 & Particle charge                \\

        $w_\mathrm{\textsc{Puppi}}$  & Weight assigned by \textsc{Puppi} algorithm                                 \\

        pid                    & One-hot particle ID (8 in total) \\

        $d_0$           & Transverse Impact Parameter            \\
        $d_z$     & Longitudinal Impact Parameter     \\ 

\toprule
        
    \end{tabular}
    \label{tab:Distilinput}
\end{table}

\section{Physics Performance}
\label{sec:pp}

\begin{figure*}[t!]
    \centering
    \includegraphics[width=0.466\textwidth]{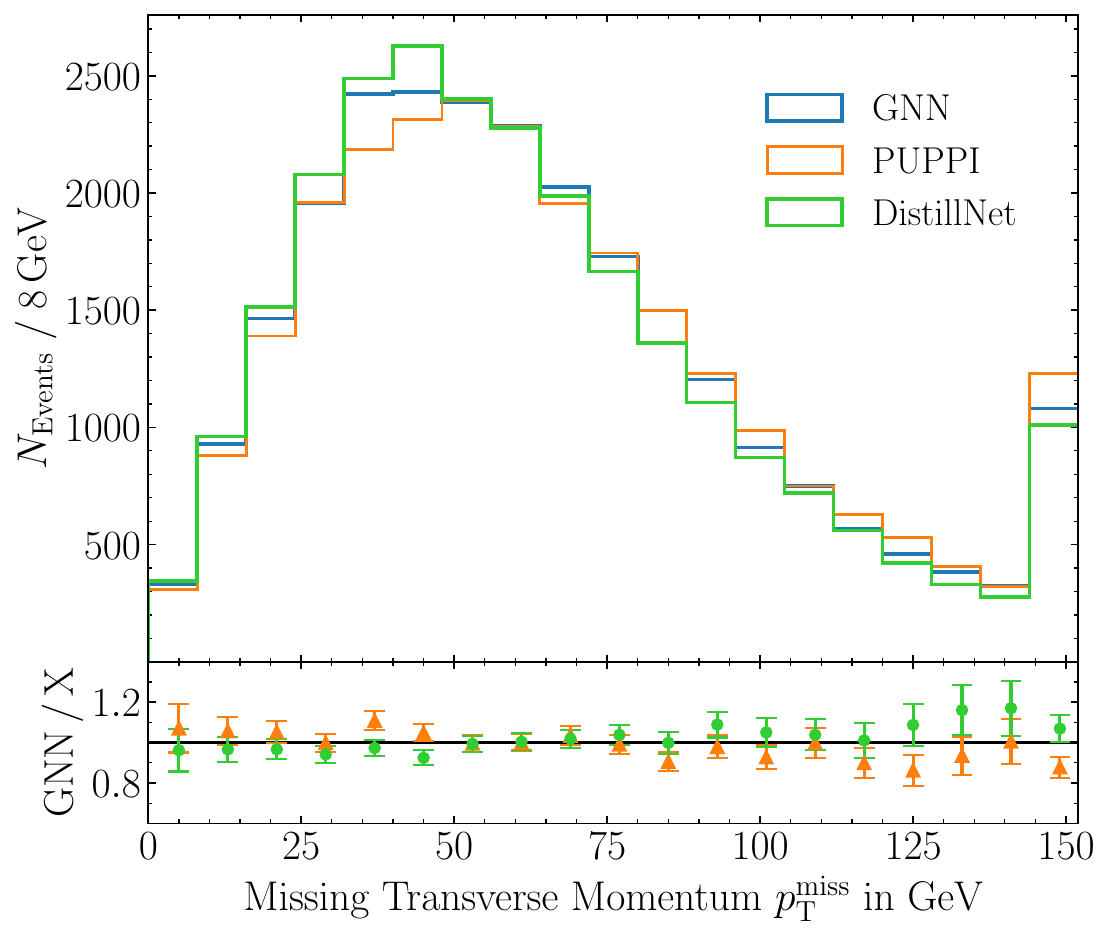}
    \includegraphics[width=0.46\textwidth]{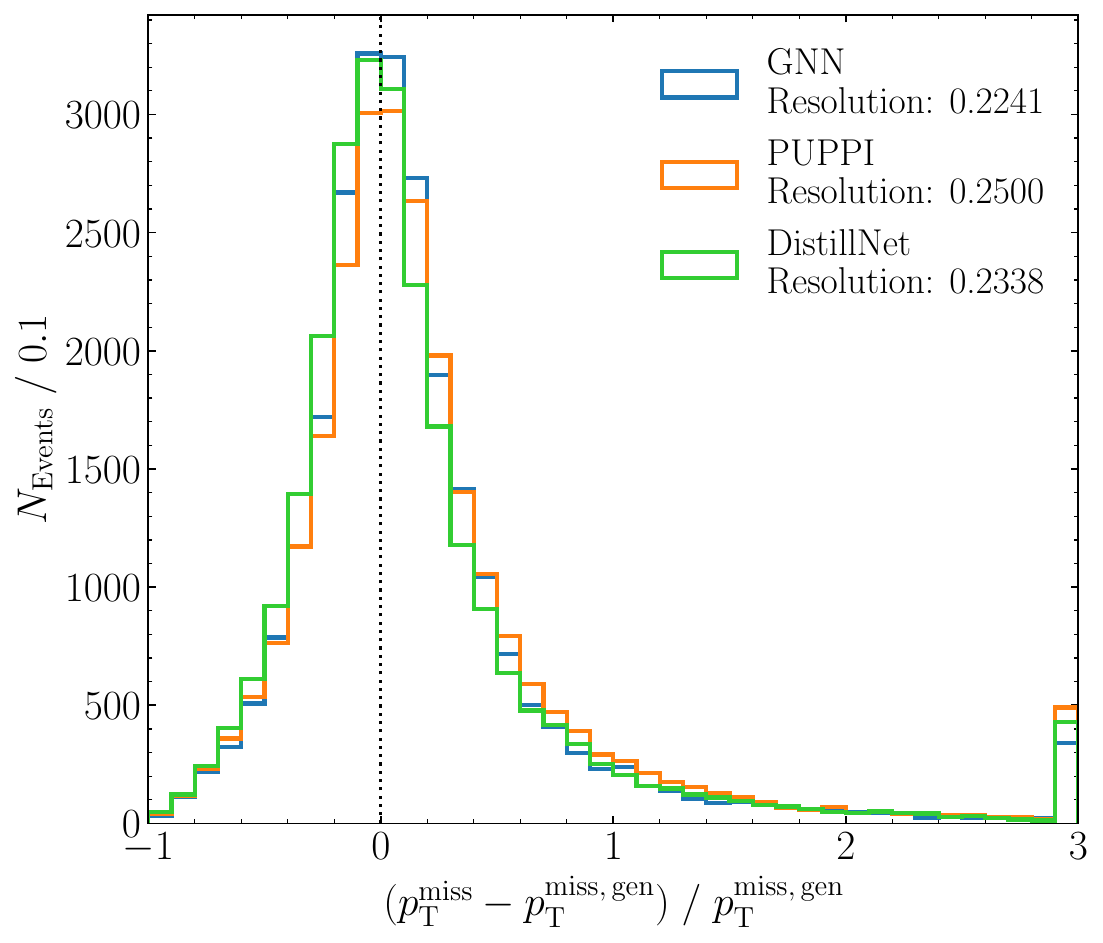}\\
    \includegraphics[width=0.472\textwidth]{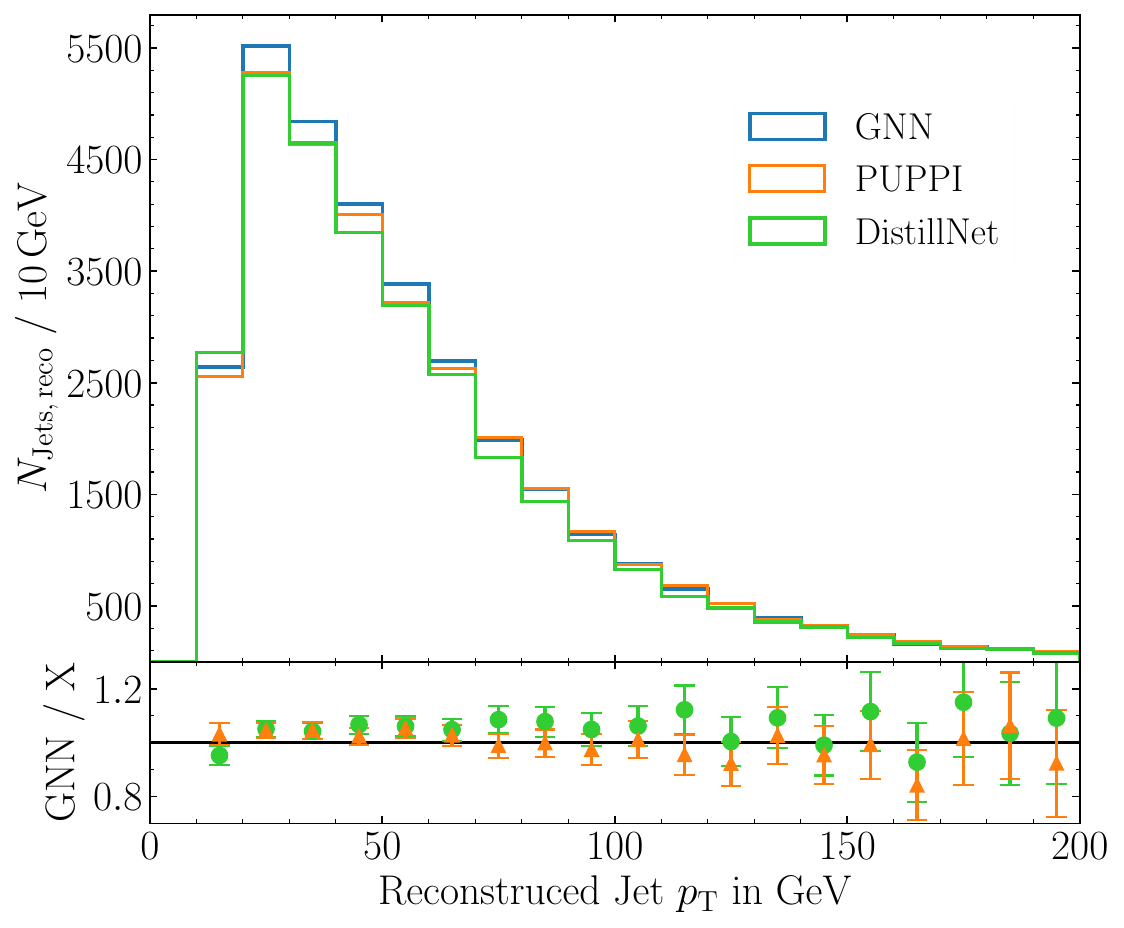}
    \includegraphics[width=0.46\textwidth]{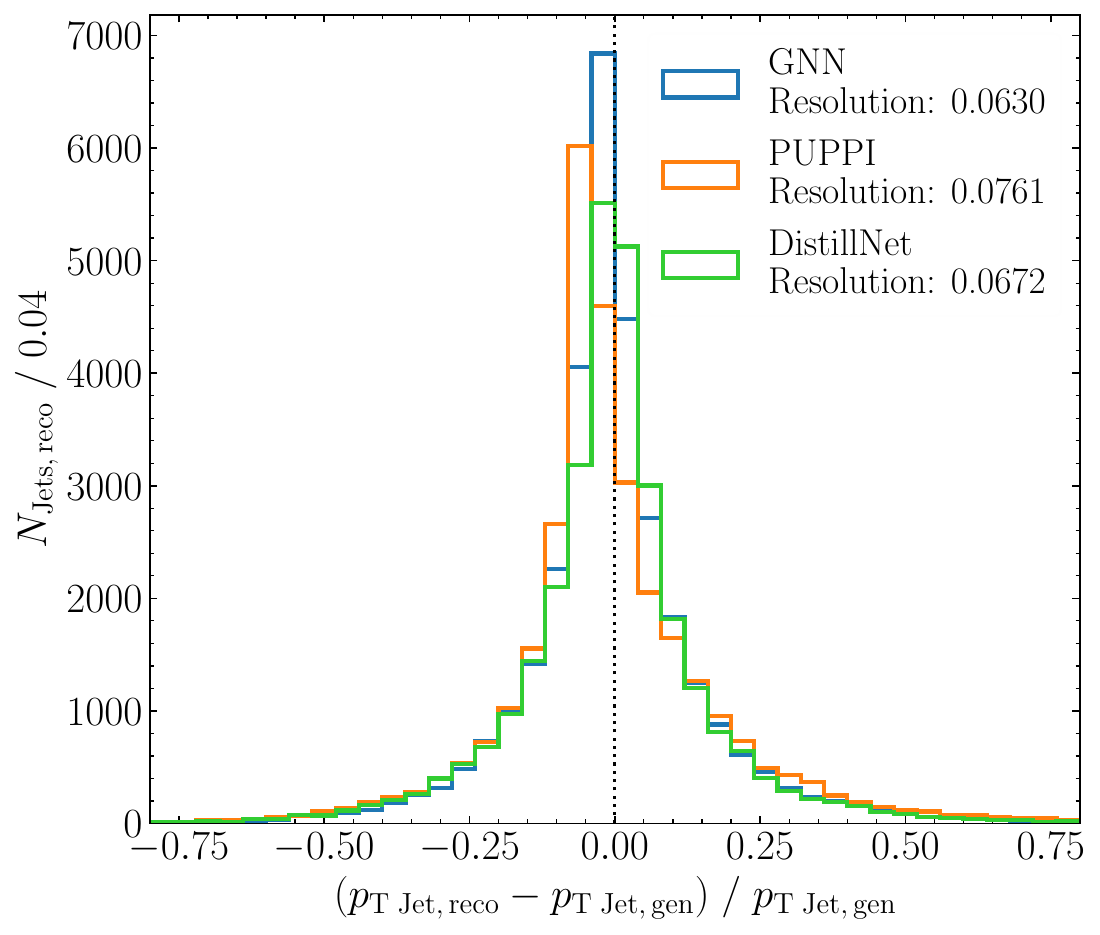}\\
    \caption{Upper row: distributions for \ptmiss (left) and the \ptmiss response (right). The resolution for the three algorithms is given in the legend of the response distributions. Bottom row: same for jet energy.}
    \label{fig:MET_results}
\end{figure*}

To quantify the loss in knowledge of the student compared to the teacher, we look at the experimental resolution 
in observables that are highly sensitive to contamination from pileup. These are the momentum missing from the event in the transverse detector plane, \ptmiss, and the transverse momentum of hadronic particle jets clustered with the anti-$k_t$ algorithm~\cite{Cacciari:2008gp} using a distance parameter of 0.4 and applying a $p_\text{T}$ cut of \SI{15}{\GeV}. In Ref.~\cite{PhysRevD.108.096003}, a sizable improvement in the resolutions for the teacher network compared to a classical pileup mitigation algorithm, \textsc{Puppi}~\cite{Bertolini:2014bba}, has been demonstrated. Additionally, the teacher yielded identical performance to another ML-based pileup mitigation algorithm, \textsc{Puma}. A loss in knowledge during the transfer from the teacher to the student network should yield less precise labels $\{y'\}$, which will translate into inferior pileup mitigation and thus a poorer experimental resolution in said quantities. We define the resolution in observable $X$ as $(q_{75\%}-q_{25\%})/2$ in its response distribution, which in turn is given by $(X-X_\text{gen})/X_\text{gen}$. Here, $q_{n\%}$, is the n-th percentile, and $X_\text{gen}$ is the observable calculated in a sample without pileup and with no detector effects.

\autoref{fig:MET_results} shows the distributions for \ptmiss and jet energy after the per-particle weights have been propagated to the \ptmiss computation and, respectively, the jet clustering. The \ptmiss shape for \DN~follows closely the GNN and overall exhibits a sound behavior, with large \ptmiss tails completely absent. Their presence would be a smoking gun for pathological events from the generation of sizable fake \ptmiss. The resolutions in \ptmiss for the teacher GNN are also presented. While the GNN yields a 10\% inclusive improvement in resolution for \ptmiss over \textsc{Puppi}, \DN~shows a 6-7\% improvement over the benchmark. For jets, the improvement over \textsc{Puppi} achieved with the GNN and with \DN~are 18\% and 12\%, respectively. This means that \DN~is able to retain most of the information contained in the GNN and thus most of the improvement over \textsc{Puppi}. At the same time, the per-particle \textsc{Puppi} weight is the most important input feature for the knowledge distillation process, as it carries the most domain knowledge for the  problem of pileup mitigation. In fact, knowledge transfer was found to be minimal when omitting this variable from the list of input features, as particle inter-relations are only represented through this feature. 

\begin{figure}[t]
    \centering
    \includegraphics[width=0.6\textwidth]{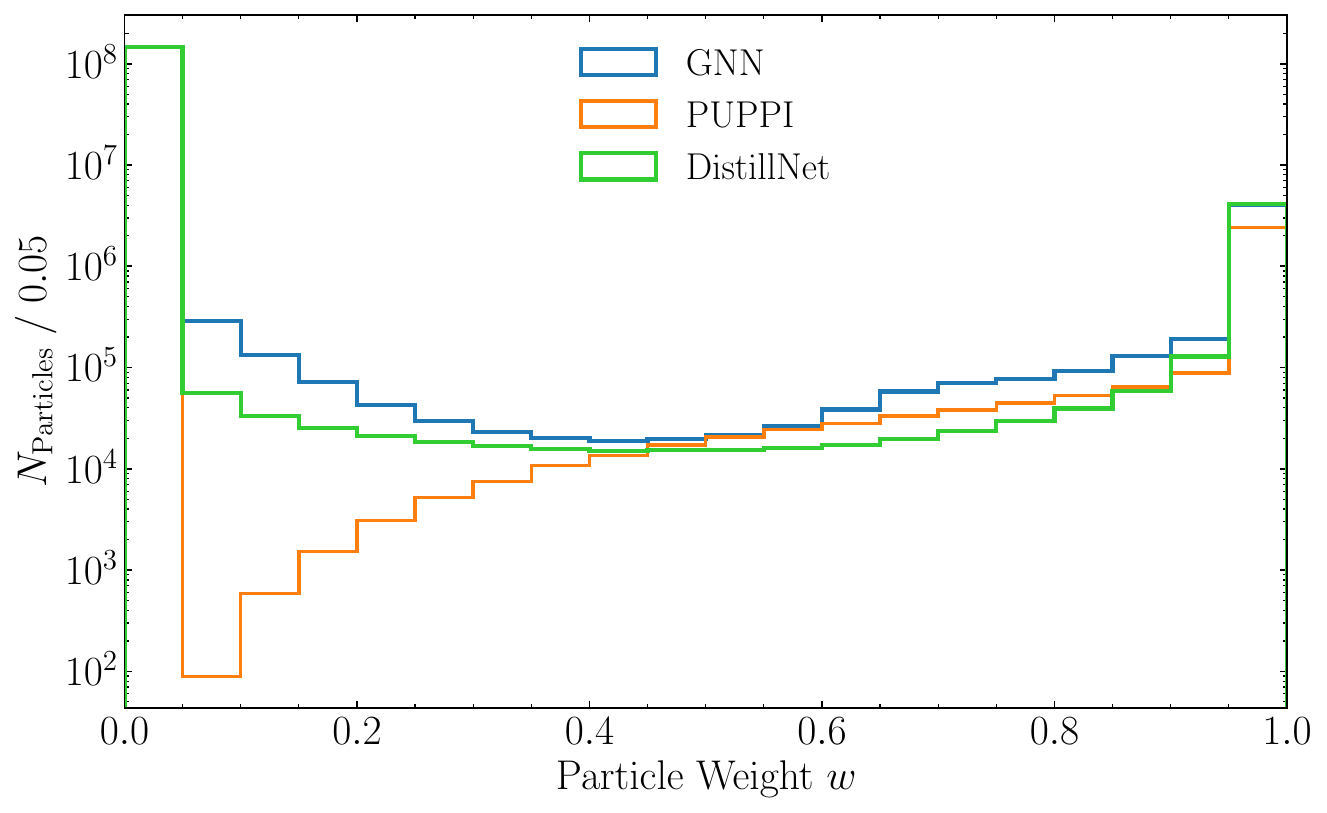}
    \caption{Per-particle weight distributions of the GNN, \DN, and \textsc{Puppi}. Particles from the hard interaction should receive a weight of $\sim$1, while pileup particles should get $w\sim0$.}
    \label{fig:total_weight_distribution}
\end{figure}

\begin{figure*}[t]
    \centering
    \includegraphics[width=0.315\textwidth]{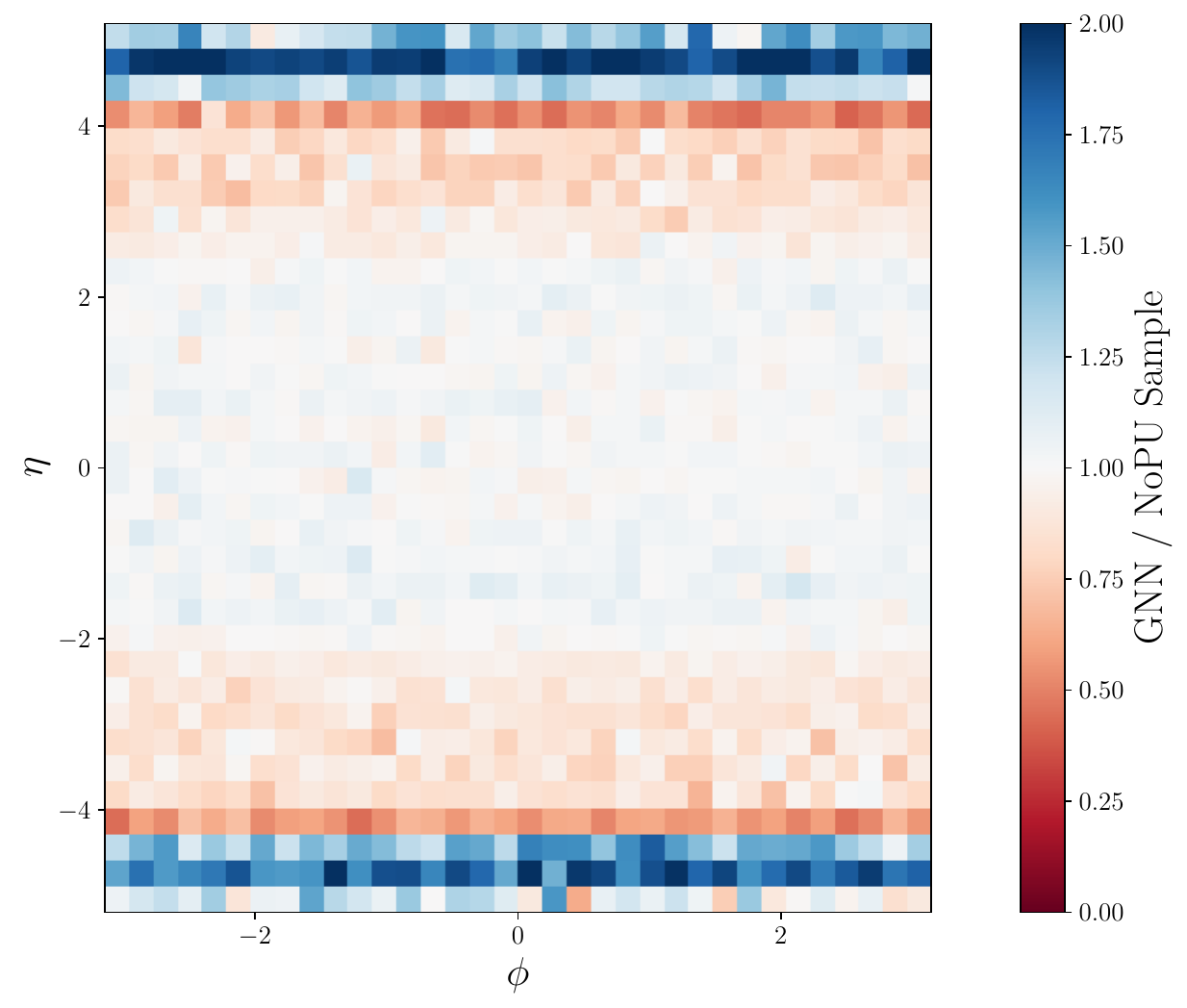}
    \includegraphics[width=0.315\textwidth]{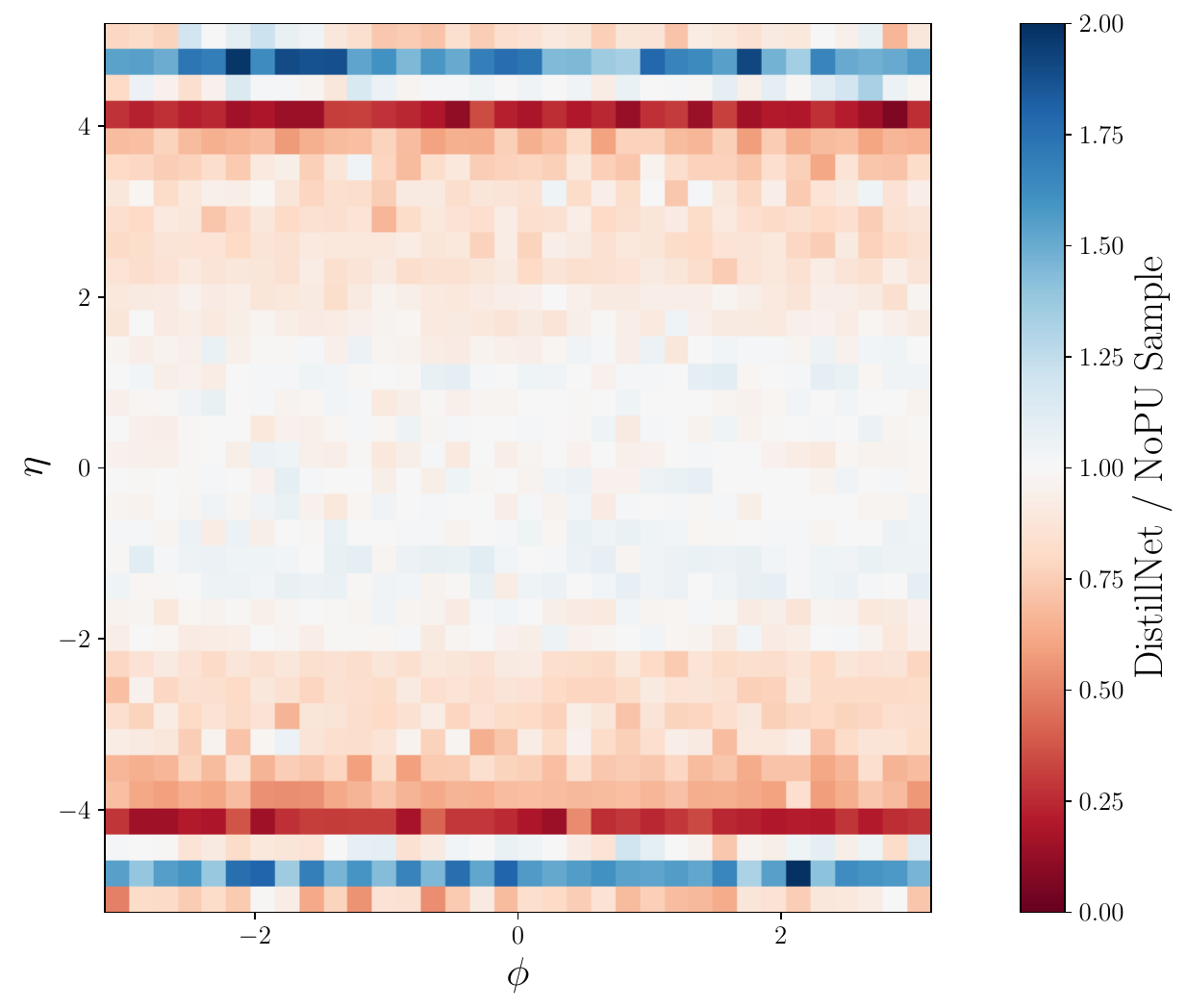}
    \includegraphics[width=0.315\textwidth]{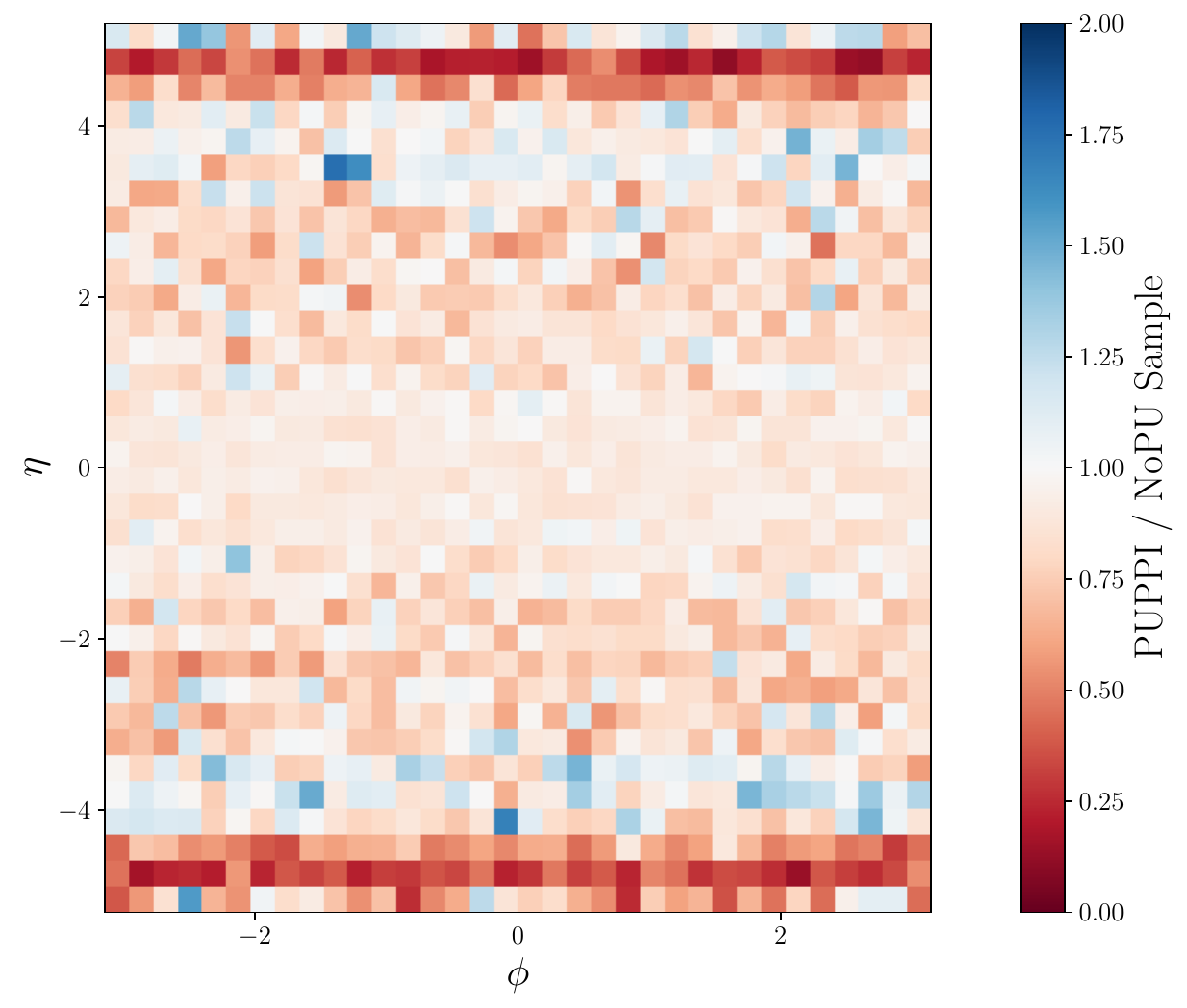}\\
    \caption{Ratio between the per particle $p_\mathrm{T}$ of the GNN, \DN, and, respectively, \textsc{Puppi} and the $p_\mathrm{T}$ of the corresponding no pileup sample in the {{$\eta\,\text{-}\,\phi$}} space of a 2D histogram. The blue bins indicate an overestimation of activity, while the red bins indicate an underestimation.}
    \label{fig:plot2d}
\end{figure*}

The similarity in observables between GNN and \DN~prompts us to compare the actual distributions of the per-particle weights for the different algorithms in \autoref{fig:total_weight_distribution}. As expected, the distributions for the teacher and student network, ${y}$ and ${y'}$ in \autoref{eq:MAE_loss} are very similar, with \DN~also being able to mimic the GNN behavior for particle candidates whose reconstruction is driven by neutral towers with merged energy deposits from pileup particles and particles from the hard interaction in the area $0<w<1$. 
We furthermore study the weights differentially in the detector $\eta$-$\phi$ plane. As shown in \autoref{fig:plot2d}, \DN~faithfully reproduces the behavior of the GNN, with problems already present in the teacher like in transition regions and at tracker edges slightly more accentuated. \textsc{Puppi} on the other hand appears to be quite aggressive in the high-$\eta$ region of the detector.
Overall, we find the performance of \DN~to be in between the two benchmarks given by the GNN and \textsc{Puppi}.

\section{Latency Studies}

An improvement over a classical benchmark as presented in \autoref{sec:pp} is one of the requirements to consider knowledge distillation. The primary motivation, however, is to 
find a compromise between the optimal physics performance provided by the teacher and significantly reduced latencies. A smaller inference time and reduced memory requirements can be useful in offline scenarios like large-scale processing and simulation campaigns, where a pragmatic choice in this two-dimensional plane can yield faster turnaround times and significantly reduced incurred costs. At the same time, an ultra-fast neural network algorithm can be deployed on FPGAs in the online selection stage for a more informed decision on whether to keep or discard a collision event. For instance, a better \ptmiss estimate will help to be more efficient at rejecting background events with fake \ptmiss and result in a higher yield for events involving electroweak processes or potential dark matter particles. To that end, we perform timing studies on a high-performance computing CPU as well as on an FPGA as foreseen for the trigger stages at the HL-LHC experiments.

\subsection{On CPU}

The CPU timing studies are performed on a single core of an Intel\textsuperscript{\tiny\textregistered} Xeon\textsuperscript{\tiny\textregistered} CPU E5-2650 v4 with an x86-64 architecture, operating with an enabled frequency boost to a maximum turbo clock speed of $\SI{2.90}{\giga\hertz}$. The timing script itself does not employ hyper-threading and only uses a single thread. When comparing the raw numbers of the networks in \autoref{tab:flops}, their size difference becomes apparent: the teacher GNN exceeds \DN~in the number of trainable parameters by a factor 27, and a forward pass consists of 35 times as many floating point operations. 
One of the key differences between the GNN and \DN~is the hierarchy at which they operate. A single sample passed to the GNN comprises an entire event, which has up to 9000 (zero-padded) reconstructed particle candidates. On the other hand, \DN~works with single particles and thus breaks the event correspondence between them. For a fair comparison, a single forward pass through the GNN therefore corresponds to 9000 forward passes for \DN, whose latencies have to be summed.

\begin{table}
\centering
\caption{Parameter comparison between teacher (GNN) and student (\DN). The number of floating point operations is provided for an event with 9000 particle candidates.}
\begin{tabular}{l|c|c}
\toprule
& Trainable Params. & MFLOPs\\ 
\hline
GNN & 270k & 6590 \\
\hline
\DN  & 10k & $0.0209\cdot 9000\sim 188$ \\
\toprule
\end{tabular}
\label{tab:flops}
\end{table}

\begin{figure}[t]
    \centering
    \includegraphics[width=0.6\textwidth]{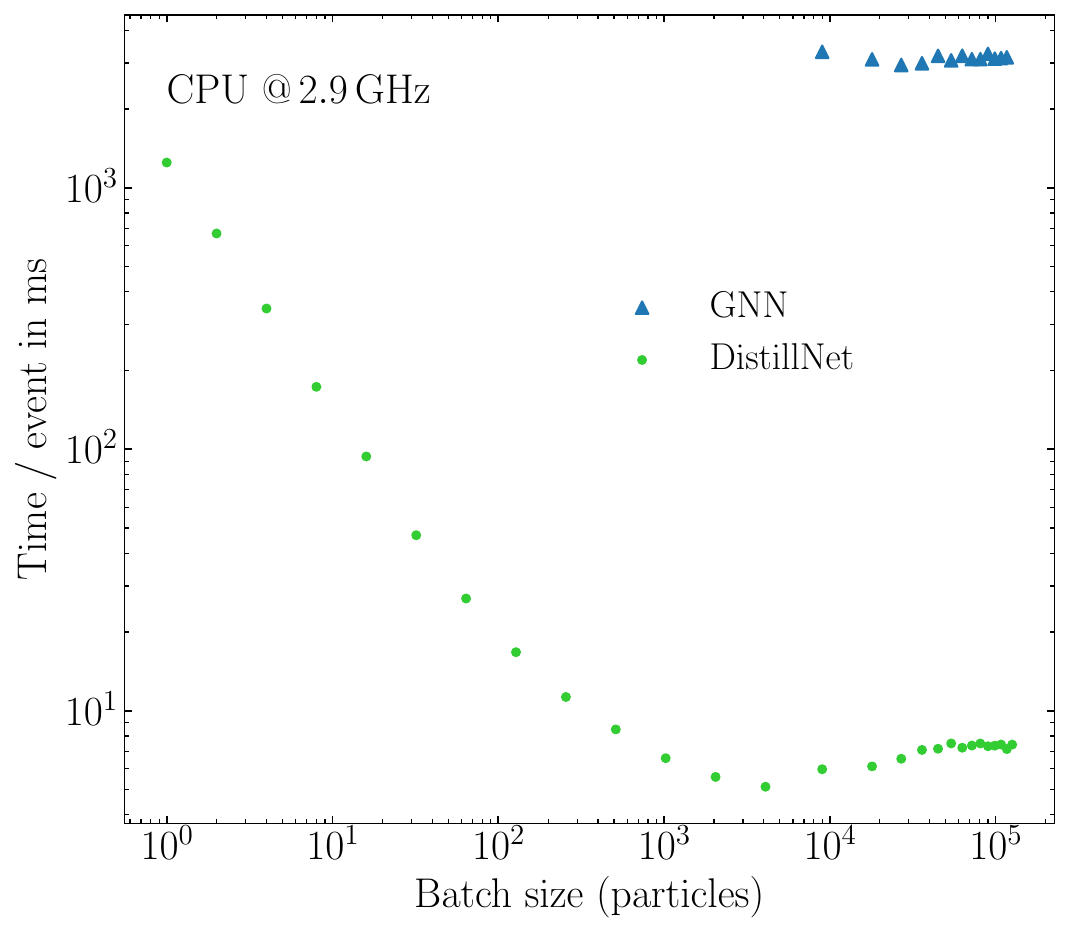}
    \caption{Per-event inference time of \DN~and the GNN as a function of batch size on CPU. }
    \label{fig:timing} 
\end{figure}

However, the vectorization capabilities of modern machine learning frameworks like PyTorch~\cite{NEURIPS2019_9015} utilizing the SIMD (Single Instruction, Multiple Data) paradigm will make operations much more efficient, thereby saving dramatically in inference time. 
This is shown in \autoref{fig:timing}. Scanning the batch size that corresponds to the number of particles, we first measure the latency of the GNN when processing the content of one event. For the GNN, since one sample comprises one event, the timing measurements correspond to multiples of 9000 particles. It can be seen that the inference time per event is about 3\,s for the GNN and stays constant for batch sizes of 1 to 10 events, i.e., of 9000 to 90000 particles\footnote{Our specific GNN implementation uses a constant number of FLOPs, as zero-padded particles are added to the input data up to a fixed-length input of 9000. In theory, the number of FLOPs and thus the evaluation time scales linearly with the number of particles}. The nature of graph-based data involves varying graph sizes, connectivity, and structure, making it challenging to effectively parallelize across the entire batch. On the other hand, \DN~exhibits remarkable performance advantages, particularly when handling larger numbers of particles per batch, with improvements in latency up to almost three orders of magnitude due to the easily vectorizable calculations executed by the deployed linear layers. Specifically, in \DN's matrix-vector product, only the matrices undergo an increase in size as the batch sizes increases. Consequently, larger matrices contribute to reduced inference times, provided they fit within the CPU cache.

The primary constraints impacting performance on the CPU are twofold: the cache size of the CPU becomes a bottleneck as batch sizes grow substantially (beyond $\mathcal{O}$(10.000) particles). Conversely, the maximum clock speed becomes a limiting factor at smaller batch sizes.


\subsection{FPGA Implementation}

Pileup removal will be a crucial component of the hardware event filtering systems of particle detectors during the HL-LHC phase. These Level-1 trigger systems, so-called will consist of $\mathcal{O}$(1000) field programmable gate arrays (FPGAs) tasked with reducing the data rate driven by the bunch collision frequency of 40\,MHz to 750\,kHz in real-time. In particular for the CMS experiment upgrade, charged tracks will be processed and reconstructed in this system~\cite{CERN-LHCC-2020-004}. This will allow for information to be combined with data from the calorimeters and muon chambers, in order to reconstruct ``Particle Flow'' (PF)~\cite{l1pf} particle candidates resembling the input we obtain from the Delphes E-Flow reconstruction. After reconstruction of the particle candidates, they are passed to a modified \textsc{Puppi} algorithm where pileup removal is performed. For simplicity reasons, we perform our latency studies with the dataset consisting of the E-Flow candidates used in the earlier section, even though a degradation in the accuracy and expressiveness of particle attributes is to be expected due to coarser information available to online algorithms. The total allocated latency to perform both PF reconstruction and \textsc{Puppi} pileup removal is 1.5\,$\upmu$s according to~\cite{CERN-LHCC-2020-004}. The \textsc{Puppi} hardware implementation has a latency of O(100\,ns). Since \textsc{Puppi} is an inherently local algorithm, it can be parallelized. It is foreseen that O(10) copies of the algorithm are running in parallel, each receiving a particle on every clock tick.

To estimate whether \DN~is suitable for pileup removal in the Level-1 trigger, the model is compressed at training time. First, the architecture is reduced to 64 and 32 neurons in each hidden layer, respectively. Secondly, the model is trained quantization-aware using QKeras~\cite{qkeras} to a bitwidth of 8, meaning each weight in the network can only take on one of $2^8=256$ discrete values. To further reduce the model resource consumption, small weights that do not contribute much to the overall decision are removed in an iterative procedure referred to as \textit{pruning}. This is achieved using the TensorFlow Model Optimization toolkit. The model is pruned to 75\% sparsity, meaning 75\% of the weights are removed from the network. The final accuracy of this compressed model is 97.7\%, and is hereby referred to as Q\DN. This presents a minor performance loss compared to \DN, which has an accuracy of 98.9\%, but still improves upon the 96.6\% accuracy of \textsc{Puppi}. In our case, we define the accuracy as the mean of the per bin ratio of predicted weights to the total number of GNN weights of the per-particle weight distribution shown in \autoref{fig:total_weight_distribution}.

This model is translated into FPGA firmware using the synthesis framework \hlsfml~\cite{Duarte:2018ite}. Due to the small network size, a fully parallel implementation is used. The target FPGA is a Xilinx Virtex Ultrascale 13+ with a clock frequency of 240\,MHz (part \textbf{xcvu13p-flga2577-2-e}). This is the same hardware and clock frequency that is foreseen for the upgraded CMS Level-1 trigger at HL-LHC. The resources at disposal on the FPGA are digital signal processors (DSPs), lookup tables (LUTs), memory (BRAM) and flip-flops (FF). The BRAM is only used as a look-up table read-only memory for calculating the final sigmoid output. 

\begin{table}[t]
    \centering \footnotesize
     \caption{Accuracy, latency and resource utilization for Q\DN. Resources are expressed as a percentage of those available on a Xilinx VU 13+ (part \textbf{xcvu13p-flga2577-2-e}).}
    \begin{tabular}{c|c|c|c|c|c|c|c}
                & Acc.   & Latency [ns] & II [cc] & BRAM & LUT           & FF             & DSP \\
        \hline
        Q\DN  & 97.7\% & 25          & 1 & 1 (0.02\%)& 2.379 (0.2\%) &  441 (0.01\%) & 6 (0.05\%)\\
    \end{tabular}
    \label{tab:fpga_res}
\end{table}

The final latency and resource utilization report for Q\DN~is shown in~\autoref{tab:fpga_res}. The latency of one single-particle inference is measured as 25\,ns, which is smaller or comparable to the foreseen implementation of the \textsc{Puppi} algorithm and could be deployed consecutively with \textsc{Puppi} in order to utilize the \textsc{Puppi} weight as a powerful input variable. The {\em initiation interval (II)}, the number of clock cycles between when the algorithm is ready to receive new data inputs, is 1. That means Q\DN~can continuously receive new particle input. The resource consumption of a single instance of Q\DN~is minimal, using ${<}0.5\%$ of the FPGA resources. It would therefore be feasible to run several of these networks in parallel at very low resource cost. Assuming ten copies of Q\DN, the total resource consumption would still be below $2\%$ of the resources available on the FPGA.

\section{Discussion}

We have presented a successful knowledge distillation (KD) from a graph neural network (GNN) to a deep neural network (DNN) to solve the challenging task of pileup mitigation at greatly reduced computational cost at the High-Luminosity LHC. Starting from a pre-trained teacher operating at the event level, the DNN, which works with particle-level input, learns to approximate the teacher output, yielding an improvement in resolution in key experimental quantities. This holds true both for an offline scenario, where large-scale data processing campaigns could profit from a reduction in latencies compared to the GNN of up to 3 orders of magnitude, as well as in real-time with an implementation of a quantized and pruned algorithm on FPGAs. Our study marks the first time that the information encoded in a graph neural network is distilled into a small deep neural network to solve reconstruction tasks at high-energy physics experiments, specifically, the problem of particle provenance in extreme environments such as at the High-Luminosity LHC. 

If implemented at LHC experiments, either in the trigger stages or for offline analysis, DistillNet will improve the selection efficiency for scenarios of physics beyond the standard model involving invisible particles and significant hadronic activity, while keeping computational costs at a minimum compared to alternative machine learning algorithms. Especially when operating at real time in the trigger, its limitation are primarily the constraints (e.g., the number of parallel processing units) of the hardware system and the degree domain-specific information is available through the input features. Future research should explore other applications of KD such as the compression of powerful jet tagging algorithms, or the extension of target architectures beyond DNNs, such as convolutional neural networks or, more generally, architectures that lie between GNNs and DNNs in terms of their complexity.

\section*{Data Availability Statement}

\sloppypar The datasets used for training and testing DistillNet are published at~\href{https://doi.org/10.5281/zenodo.10670352}{https://doi.org/10.5281/zenodo.10670352}. The training code is available at~\href{https://github.com/tbrandes01/DistillNet}{https://github.com/tbrandes01/DistillNet}.

\section*{Acknowledgments}

BM acknowledges the support of the Alexander von Humboldt Foundation and of Schmidt Futures. FI is supported by the Ministry of Science and Technology of China, project No. 2018YFA0403901 and National Natural Science Foundation of China, project No. 12188102, 12061141003. T\AA~is supported by the Swiss National Science Foundation Grant No.~PZ00P2\_201594. This research used resources of the National Energy Research Scientific Computing Center, a DOE Office of Science User Facility supported by the Office of Science of the U.S. Department of Energy under Contract No. DE-AC02-05CH11231 using NERSC award HEP-ERCAP0021099.


\bibliography{references}
\end{document}